\documentclass[twocolumn,aps,amssymb,amsmath,eqsecnum,nofootinbib,floatfix]{revtex4}
\begin{document}

\title{Second order gauge invariant measure of a tidally deformed black hole}

\author{Nahid Ahmadi\footnote{~nahmadi@ut.ac.ir}}

\affiliation{Department of Physics, University of Tehran, Kargar Avenue North, Tehran 14395-547, Iran}

\begin{abstract}
In this paper, a Lagrangian perturbation theory for the second order treatment of small disturbances of the event horizon in Schwarzchild black holes is introduced. The issue of gauge invariance in the context of general relativistic theory is also discussed. The developments of this paper is a logical continuation of the calculations presented in \cite{Vega+Poisson}, in which the first order coordinate dependance of the intrinsic and exterinsic geometry of the horizon is examined and the first order gauge invariance of the intrinsic geometry of the horizon is shown. In context of second order perturbation theory, It is shown that the rate of the expansion of the congruence of the horizon generators is invariant under a second order reparametrization; so it can be considered as a measure of tidal perturbation. A generally non-vanishing expression for this observable, which accomodates tidal perturbations and implies nonlinear response of the horizon, is also presented.
\end{abstract} 
\maketitle
\def\be{\begin{equation}}
\def\ee{\end{equation}}
\def\bea{\begin{eqnarray}}
\def\eea{\end{eqnarray}}
\def\nn{\nonumber}
\newcommand{\bes}{\begin{subequations}}
\newcommand{\ees}{\end{subequations}}

\section{Introduction and Motivation}

The subject of evolution of compact binaries (involving neutron stars and/or black holes), as the source of gravitational radiation, have received the attention of many authors. One point which has been vigorously focused is the calculation of tidal coupling on the gravitational waves \cite{Flanagan+Hinderer, Damour and Nagar}. Finding a nonvanishing measure of tidal deformation of each body is the issue that deserves an analysis of its own. In these studies, the comparison of Einstein's theory with the gravitational wave emitted by such systems is almost entirely based on approximation methods. Also, the nature of relativistic perturbation theory is tied with gauge issues. When using the perturbational calculations to measure the tidal deformations, one must deal with the freedom to redefine the coordinates employed to describe the spacetime background. This can be tackled in two very different ways: First, by choosing a convenient gauge which simplifies the calculations for all orders of perturbation and then relating the result to an asymptotically flat gauge in which the information about the tidal signatures can be extracted. A second way is to find a coordinate independent measure for studying the tidal interactions.

The relativistic Love numbers are dimensionless parameter measures that relate the induced mass moments of a compact body to the applied tidal field. It has been shown \cite{Damour and Nagar} that these gauge invariant numbers can be detected by Earth-based gravitational detectors gathering information about tidal deformations. For nonrotating black holes, however, these numbers are equal to zero and the description of a tidally deformed black hole must be pursued in another way. Recently Vega, Poisson and Massey [1] studied the coordinate dependence of Schwarzchild event horizon quantities and showed that the intrinsic geometry of the event horizon is invariant under a reparametrization of its null generators. They considered the event horizon perturbed by a small linear disturbance and found a group of invariant quantities defined on the horizon. Although calculations to first order in some expansion parameters have proven to give simple results reliable enough to describe the tide raised during a close encounter between a black hole and a small orbiting body, in principle first order perturbation theory has a limit of applicability and when perturbations become large enough, the predictions of first order differs significantly from the higher order predictions. Furthermore, in view of the increased sensitivity expected from the next generation of detectors, second order calculations may prove necessary to have more reliable results, where linear perturbation theory is not capable of justifying the physical phenomena. In this paper we try to discuss about one phenomena that does not show up at first order, but potentially can be used to in interpretation of the details of the processes like colliding black holes as the main source of gravitational waves. 

The work of Vega et al. in \cite{Vega+Poisson} motivated an effort to improve our understanding of a tidally deformed black hole by calculating some horizon quantities to second order. Following \cite{Vega+Poisson}, we assume that i) the unperturbed black hole is nonrotating and is described by the Schwarzchild solution, ii) the tidal field can be either static or slowly varying or even fully dynamic, but it is such that the horizon differs slightly from the horizon in the unperturbed spacetime. Under these assumpsions, the instability is of no concern. At the same time it is assumed that the perturbations are large enough that the predictions of first order calculations differ significantly from that of the second order.

In practice, we describe the deformation of the event horizon by the deformation of the congruence of its null generators. Any point on the horizon can be uniquely identified with coordinates $\left(\lambda,\theta^A\right)$, where $\theta^A$ and $\lambda$ are respectively, the comoving coordinate and the running parameter on the generator that passes through the given point. The horizon quantities are a collection of fields expressed entirely in terms of the coordinates $\left(\lambda,\theta^A\right)$. Based on the calculations in \cite{Vega+Poisson}, not all of these quantities are invariant under the transformation of background spacetime, or identification gauge invariant (igi) in the language of \cite{Stewart and Walker}. In \cite{Bruni and Sonego}, Bruni and Sonego discussed the issue of observability in general relativistic perturbation theory and showed that the perturbation of a {\it scalar} field $Q$ is observable iff its representation on the background is first order igi even when it is gauge dependent at higher orders. 
This means that although the background spacetime is merely a mathematical artifice and the measurements are performed in the perturbed spacetime, as far as the observability or, as in our case, finding a measure of tidal deformation is concerned, the invariance under the transformations of background spacetime is applicable.

In the next section we review the relativistic perturbation theory of higher orders. The geometry of a nonrotating black hole perturbed by a distribution of matter either flowing across the event horizon or situated outside the black hole's immediate neighborhood is discussed in section III. In section IV, the induced transverse metric and the geometry evolution on the black hole horizon is computed to the second order in perturbation theory. Some concluding remarks will be given in section V. 

\section{Perturbations of spacetimes}

A perturbative approach in general relativity always deals with two spacetimes. The physical (perturbed) spacetime $\left(M',g'\right)$ and the mathematical (unperturbed) spacetime $\left(M,g\right)$. These two spacetimes differ only slightly from each other. The points in $\left(M',g'\right)$ and $\left(M,g\right)$ are unrelated. To define the perturbation of a quantity it is necessary to identify the points corresponding to the same physical event. Mathematically, it is convenient to treat small perturbations in a frame called "Lagrangian frame". One usually defines a Lagrangian displacement vector (and correspondingly a Lagrange gauge) connecting the configuration in unperturbed spacetime to the corresponding elements in the perturbed one.
The "Lagrange change" in a tensor, denoted by $\Delta Q$, measures the change in the components of a tensor with respect to the frame which is embedded in the unperturbed spacetime and dragged along with it by the perturbation.\footnote{A Lagrangian displacement $\xi^\alpha$ vector uniquely determines a Lagrangian change but the reverse is not true. In fact, any perturbation can be characterized by {\it some} displacement provided that it conforms to any natural restrictive conditions (or conservation laws) in the problem. However, there is a class of {\it trivial} displacements for which the Eulerian changes in physical variables all vanish. Two displacements that differ by a trivial displacement $\eta^\alpha$ are the same. In other words, for a physical variable $Q$, we have $\varphi^{*}Q=\psi^{*}Q$, if $\varphi$ and $\psi$ are generated by $\xi^\alpha$ and $\xi^\alpha+\eta^\alpha$, respectively. A trivial displacement generates a pure gauge transformation in physical variables; so the background Killing vector fields are in the class of trivials \cite{Schutz and Sorkin, Friedman and Schutz}.}
 Accordingly, the geometry of the perturbed spacetime can be described by an exponential map
\be
x^{\alpha}\rightarrow e^{{\cal L}_\xi} x^{\alpha},
\ee
in which ${\cal L}_\xi$ denotes the Lie derivative with respect to the Lagrangian displacement vector, $\xi$. Up to second order in perturbation theory, $\xi$ is given by
\be
\xi^{\alpha}=\lambda\xi^{\alpha}_{\left(1\right)}+ \frac{1}{2}\lambda^{2}\xi^{\alpha}_{\left(2\right)}
,\ee
and the exponential map is then given by
\be
e^{{\cal L}_\xi}=1+\lambda{\cal L}_{\xi_{\left(1\right)}}+\frac{1}{2}\lambda^{2}{\cal L}^{2}_{\xi_{\left(1\right)}}+\frac{1}{2}\lambda^{2}{\cal L}_{\xi_{\left(2\right)}}
.\ee
Therefore, the first and second order Lagrangian changes, $\Delta Q$, in a tensor $Q$ is related to Eulerian changes, $\delta Q$, through
\bes
\bea
\Delta^{\left(1\right)}Q&=&\delta^{\left(1\right)} Q+{\cal L}_{\xi_{\left(1\right)}}Q,\\
\Delta^{\left(2\right)}Q&=&\delta^{\left(2\right)}Q+2{\cal L}_{\xi_{\left(1\right)}}{\delta}^{\left(1\right)}Q+{\cal L}^{2}_{\xi_{\left(1\right)}}Q+{\cal L}_{\xi_{\left(2\right)}}Q.\nn\\
\eea
\label{changed coordinate}\ees
Here $\xi_{\left(1\right)}$ and $\xi_{\left(2\right)}$ are the generators of the first and second order perturbations. Equation (\ref{changed coordinate}) can be applied to the coordinate function $x^\alpha$ on $M$. We have then
\be
\tilde{x}^\alpha=x^\alpha+\lambda\xi_{\left(1\right)}^\alpha+\frac{\lambda^2}{2}\left(\xi_{\left(1\right)}^\beta\xi_{\left(1\right);\beta}^{\alpha}+\xi_{\left(2\right)}^{\alpha}\right),
\ee
where the coordinate corresponding to $q=\varphi\left(p\right)$, $\varphi:M\rightarrow M'$, is denoted by $\tilde{x}^\alpha$. One can think of the Lagrange change as an \textit{active} gauge transformation on the background; so $\xi_{\left(1\right)}$ and $\xi_{\left(1\right)}$ can be considered as two independent generators of the gauge transformation map. According to the relation between the perturbations in different gauges, it is possible to define the gauge invariant quantities. A tensor field $Q$ is {\it gauge invariant} to order $n$, iff ${\cal L}_{\xi}\delta^{\left(k\right)}Q=0$, for any vector field $\xi^\mu$ defined on the space of spacetimes,${\cal M}$, and $\forall k<n$, where $\delta^{\left(0\right)}Q=Q^{\left(0\right)}$. The equations (\ref{changed coordinate}) show that it is true for order $n=2$ and the generalization to an arbitrary order $n$ can be proved by induction. 

The necessary and sufficient conditions on $Q$ to be a first order gauge invariant is discussed in \cite{Stewart and Walker}. In their terminology, the perturbation of $Q$ is igi, iff one of the following holds: i) $Q$ vanishes, ii) $Q$ is a constant scalar, iii) $Q$  is a constant linear combination of products of Kronecker deltas. The fact that the intrinsic geometry of the event horizon and the curvature tensor is gauge invariant can be seen from these conditions .

Let $Q$ be a first order gauge invariant tensor; so ${\cal L}_{\xi_{\left(1\right)}}Q=0$, $\forall\xi$. If the first order change in $Q$ be also zero, we will have $\tilde{\delta}^{\left(1\right)}Q=0$ and accordingly ${\cal L}_{\xi_{\left(1\right)}}{\delta}^{\left(1\right)}Q=0$. This situation provides us a second order gauge invariant tensor. Among the first order gauge invariant quantities discussed in \cite{Vega+Poisson}, the rate of expansion of the congruence of null generators denoted by $\Theta$, vanishes to leading order in perturbation theory; therefore $\Theta$ is a second order gauge invariant scalar. Furthermore, $\Theta$ is a scalar field and we have $\varphi_*\Theta=\psi_*\Theta$, $\forall \varphi,\psi:M\rightarrow M'$, which means that its value at any point of ${M}'$ does not depend on the gauge choice and describes an observable. For the remaining first order invariant quantities, this conclusion may not be valid. This characteristics in $\Theta$ motivated us to evaluate it up to second order in the next section.
\section{tidally deformed black hole}

We consider a nonrotating black hole as the unperturbed spacetime and describe its metric by Schwarzchild's solution described in Eddington-Finklestein coordinates,
\be
g^{\left(0\right)}_{\alpha\beta}dx^\alpha dx^\beta=-f dv^2+2dv dr+r^2\Omega_{AB} d\theta^A d\theta^B
,\ee
where $f:=1-\frac{2M}{r}$ and $\Omega_{AB} d\theta^A d\theta^B:= d\theta^2+\sin{^2}\theta d\phi^2$ is the metric on the unit two sphere 
with the inverse denoted by $\Omega^{AB}$. In this paper, upper case Roman letters $A,B,C \ldots$ are used for the indices on two dimensional unit sphere; covariant derivative compatible with $\Omega_{AB}$ is denoted by $D_A$. The event horizon in the unperturbed spacetime can geometrically be described by null geodesics that generate the hypersurface, each labled by $\alpha^A=\theta^A$ and a running parameter $\lambda=v$ distinguishes the points on the generator. 
\subsection *{Perturbed horizon}
The complete metric of the perturbed spacetime is $g_{\alpha\beta}=g^{\left(0\right)}_{\alpha\beta}+p_{\alpha\beta}+\frac{1}{2}p_{\alpha\lambda}p^{\lambda}_{\ \beta}$. We are interested in a Lagrangian displacement vector which determines the horizon perturbation as follows: Let the intrinsic coordinates on the perturbed horizon be $\left(v,\alpha^A\right)$ and $k^\alpha$ be the vector field tangent to the perturbed horizon generators, defined by $k^\alpha=\left(\frac{\partial\tilde{x}^\alpha}{\partial v}\right)_{\alpha^A}$, where $\tilde{x}^\alpha$ are the coordinates of the points on the perturbed horizon from the space of horizons, ${\cal M}$, point of view. The displacement vectors orthogonal to $k^\alpha$ that point from one generator to another is $e^{\alpha}_{A}=\left(\frac{\partial\tilde{x}^\alpha}{\partial \alpha^A}\right)_v$. 
We suppose that there is a diffeomorphism $\varphi_\lambda$ from the the support in unperturbed horizon $\varphi_0\left(p\right)$ to its support in $\varphi_\lambda\left(p\right)$ such that the Eulerian change (the change at fixed coordinate values, intrinsic coordinates for example) in $k^\alpha$ be the result of its being dragged along by the perturbation, i.e. 
\be
\delta k^{\alpha}=-{\cal L}_\xi k^{\alpha}\label{constraint on k}.
\ee
It also means that
\be
\Delta k^{\alpha}=\Delta e^{\alpha}_A=0\label{constraint on k and e}.
\ee
In this way, the the old $k^\alpha=\delta^{\alpha}_v$ would be tangent to the perturbed horizon, even when the spacelike surface spanned by $e^\alpha_A$ is not a marginally outer trapped surface. If we describe the conditions on the horizon by $\delta k^\alpha=\delta e^\alpha_A=0$, the perturbation will be described in horizon-locking gauge and it takes horizon generators to horizon generators. In our case, however, the perturbed horizon generators are given by
\be
\Delta\left(k^\alpha k_\alpha\right)=0,\ \ \ \ \Delta\left(e^{\alpha}_A k_\alpha\right)=0,\label{basis vectors1}
\ee
and
\be
\Delta\left(N^\alpha N_\alpha\right)=0,\ \ \ \ \Delta\left(N^\alpha k_\alpha\right)=0,\ \ \ \ \Delta\left(N_\alpha e^{\alpha}_A\right)=0,\label{basis vectors2}\ee
where $N^\alpha$ is the transverse vector needed to complete the basis vectors on the horizon. The first relation in (\ref{basis vectors1}) implies that 
\bes
\bea
\delta^{\left(1\right)}\left(k^\alpha k_\alpha\right)&=&-{\cal L}_{\xi_{\left(1\right)}}\left(k^\alpha k_\alpha\right),\label{constraint kk1}\\
\delta^{\left(2\right)}\left(k^\alpha k_\alpha\right)&=&\left[-{\cal L}_{\xi_{\left(2\right)}}-2{\cal L}_{\xi_{\left(1\right)}}{\delta}^{\left(1\right)}-{\cal L}^{2}_{\xi_{\left(1\right)}}\right]\left(k^\alpha k_\alpha\right)\nn\\
&=&-{\cal L}_{\xi_{\left(2\right)}}\left(k^\alpha k_\alpha\right)+{\cal L}^{2}_{\xi_{\left(1\right)}}\left(k^\alpha k_\alpha\right).
\eea
\label{constraint on kk}\ees
The equation (\ref{constraint on kk}) vanishes for all $\xi^\alpha$ iff $k^\alpha k_\alpha=0$ which is valid in the unperturbed system. We do not need to assume $k^\alpha k_\alpha=0$ to describe the horizon. We permit an Eulerian change in $k^\alpha k_\alpha$ as the result of its being dragged along by $\xi^\alpha$. The equation (\ref{constraint kk1}) can be expressed in the form
\be
p_{vv}=\left(k^\alpha\xi_{\left(1\right)\alpha}\right)_{;\beta}k^\beta-k^\alpha_{;\beta}k^\beta\xi_{\left(1\right)\alpha},
\ee
so we have
\be
\partial_{v}\left(k^\alpha\xi_{\left(1\right)\alpha}\right)=p_{vv}+\kappa\left(k^\alpha\xi_{\left(1\right)\alpha}\right),
\ee
where $\kappa$ is the black hole surface gravity. This calculation shows that even if the event horizon starts in a stationary state, which means $\xi^\alpha=0$ initially, $k^\alpha\xi_{\left(1\right)\alpha}$ will be nonzero until the perturbation is switched off. We therefore have $\xi^r\neq0$ (as opposed to the situation discussed in horizon-locking gauge; see for example, Sec VI of \cite{Poisson}). Furthermore, $\xi^\alpha$ proportional to $\delta^\alpha_v$ is a trivial displacement vector field, recalling the fact that it is the background Killing vector.

For the linear perturbation of the horizon generated by the vector field $\xi_{\left(1\right)}^{\alpha}=\left(0,\xi_{\left(1\right)}^r,\xi_{\left(1\right)}^A\right)$, from the relations in (\ref{basis vectors1}) together with the conditions on the displacement vector (\ref{constraint on k}) and (\ref{constraint on k and e}), we will get
\bea
0=\Delta ^{\left(1\right)}\left(k^\alpha k_\alpha\right)&=& \Delta^{\left(1\right)}g_{vv}=\delta^{\left(1\right)}g_{vv}+{\cal L}_{\xi_{\left(1\right)}}g_{vv}\nn\\
&=&p_{vv}-\left(2M\right)^{-1}\xi_{\left(1\right)}^r+2\dot{\xi}^{r}_{\left(1\right)},\label{first order kk}\\
0=\Delta ^{\left(1\right)}\left(k_\alpha e^{\alpha}_A\right)&=& \Delta^{\left(1\right)}g_{v A}=\delta^{\left(1\right)}g_{v A}+{\cal L}_{\xi_{\left(1\right)}}g_{vA}\nn\\
&=&p_{vA}+\dot{\xi}_{\left(1\right)A}+D_{A}{\xi}_{\left(1\right)}^r.\nn\\
\label{first order ke}\eea
Here an overdot indicates differentiating with respect to $v$ and $\xi_A=\left(2M\right)^2\Omega_{AB}\xi^B$ is the covariant vector corresponding to the displacement vector defined on the horizon. In this paper, we use the definitions $D_A \xi^r=\partial_A\xi^r$, $D_{A} p_{vv}=\partial_A p_{vv}$ and $D_{B}p_{vA}=\partial_Bp_{vA}-\Gamma^C_{AB}p_{vC}$.

In solving the perturbation equations, we assume that a horizon which starts in an initial Schwarzchild state (when $\xi=0$), is perturbed for some time by an external process. The final situation is described by another Schwarzchild horizon. This requires the vanishing of all horizon perturbations as $v\rightarrow\infty$. Assuming that the perturbation decays sufficiently fast, a {\it teleological} solution given by
\bes\bea
\xi_{\left(1\right)}^r&=&\frac{1}{2}\int_{v}^{\infty}e^{-\frac{1}{4M}\left({v}'-v\right)}p_{vv}d{v}',\\
\xi_{\left(1\right)}^A&=&\left(2M\right)^{-2}\Omega^{AB}\int_{v}^{\infty}\left[D_B\xi_{\left(1\right)}^r+p_{vB}\right]d{v}',\nn\\
\eea\label{first order xi}\ees
is required for the first order displacement vector components. It is understood that the fields on the integrand are evaluated at $v={v}',\ r=2M,\ \alpha^A=\theta^A$. 

Let the second order displacement vector, $\xi_{\left(2\right)}^{\alpha}$, be defined by its components $\left(\xi_{\left(2\right)}^v,\xi_{\left(2\right)}^r,\xi_{\left(2\right)}^A\right)$\footnote{Generally speaking, $\xi^\alpha_{\left(2\right)}=\delta^\alpha_v$ is not a Killing vector for the perturbed horizon; so we assumed a nonzero $\xi_{\left(2\right)}^v$. The constraints on perturbations leaves $\xi_{\left(2\right)}^v$ arbitrary.}. The second order change in horizon equations is then governed by
\bea
\Delta ^{\left(2\right)}\left(k^\alpha k_\alpha\right)&=&\Delta^{\left(2\right)}g_{vv}\nn\\
&=&p_{v\alpha}p^{\alpha}_{\ v}+{\cal L}_{\xi_{\left(2\right)}}g_{vv}+{\cal L}_{\xi_{\left(1\right)}}p_{vv}=0.\nn\\
\label{second order kk}\eea
Pure second order change would look exactly like (\ref{first order kk}), replacing $\xi_{\left(1\right)}^{\alpha}$ by $\xi_{\left(2\right)}^{\alpha}$. Therefore (\ref{second order kk}) will result in the following equation.
\bea
p_{v\alpha}p^{\alpha}_{\ v}-\left(2M\right)^{-1}\xi_{\left(2\right)}^r+2\dot{\xi}^{r}_{\left(2\right)}+\xi_{\left(1\right)}^{r}\partial_{r}p_{vv}\nn\\
+\xi_{\left(1\right)}^{A}D_{A}p_{vv}+2p_{vr}\dot{\xi}^r_{\left(1\right)}+2p_{vA}\dot{\xi}^A_{\left(1\right)}=0.
\label{dot xi r2}\eea
Similarly, from $\Delta ^{\left(2\right)}\left(k_\alpha e^{\alpha}_A\right)$=0, one obtains
\bea
&&p_{v\alpha}p^{\alpha}_{\ A}+\dot{\xi}_{\left(2\right)A}+\partial_{A}{\xi}_{\left(2\right)}^r+\xi_{\left(1\right)}^{r}\partial_{r}p_{vA}\nn\\
&+&\xi_{\left(1\right)}^{B}D_{B}p_{vA}+p_{v\alpha}D_A\xi^\alpha+p_{A\alpha}\dot{\xi}^\alpha=0.
\label{dot xi A2}\eea
The teleological solutions for the components of $\xi^\alpha_{\left(2\right)}$ are then given by
\bes\bea
\xi_{\left(2\right)}^r&=&\frac{1}{2}\int_{v}^{\infty}e^{-\frac{1}{4M}\left({v}'-v\right)}\nn\\
&&\left[p_{v\alpha}p^{\alpha}_{\ v}+\xi_{\left(1\right)}^{r}\partial_{r}p_{vv}+\xi_{\left(1\right)}^{A}D_{A}p_{vv}+2p_{v\alpha}\dot{\xi}^\alpha_{\left(1\right)}\right]d{v}',\nn\\
\eea
\bea
\xi_{\left(2\right)}^A&=&\left(2M\right)^{-2}\Omega^{AB}
\int_{v}^{\infty}\left[p_{v\alpha}p^{\alpha}_{\ B}+D_{B}{\xi}_{\left(2\right)}^r
\right.\nn\\
&+&\left.\xi_{\left(1\right)}^{r}\partial_{r}p_{vB}+\xi_{\left(1\right)}^{C}D_{C}p_{vB}+p_{v\alpha}D_A\xi^\alpha+p_{A\alpha}\dot{\xi}^\alpha\right]d{v}'.\nn\\
\eea\label{second order xi}\ees
The horizon equations including the auxiliary basis $N^\alpha$, can be used to find the components of this vector field. For example for $N^r$, the second relation in (\ref{basis vectors2}) yields
\bes\bea
\Delta ^{\left(1\right)}\left(k_\alpha N^{\alpha}\right)&=&\Delta ^{\left(1\right)}N^r-\Delta ^{\left(1\right)}g_{rv}\nn \\
&=&\Delta ^{\left(1\right)}N^r-p_{rv}=0,
\eea
\bea
&&\Delta ^{\left(2\right)}\left(k_\alpha N^{\alpha}\right)=\Delta ^{\left(2\right)}N^r-\Delta ^{\left(2\right)}g_{rv}+2\Delta ^{\left(1\right)}N^\alpha\Delta ^{\left(1\right)}g_{\alpha v}\nn \\
&=&\Delta ^{\left(2\right)}N^r-p_{r\alpha}p^{\alpha}_{\ v}-2{\cal L}_{\xi_{\left(1\right)}} p_{rv}+2p_{rv}p_{rv}=0.\nn\\
\eea\ees

After some straightforward calculations, our results up to second order perturbation are then given by
\bea
N^v&=&\frac{1}{2}\left[p_{rr}+\frac{1}{2}p_{r\alpha}p^{\alpha}_{\ r}+\xi^\alpha\partial_\alpha p_{rr}\right.\nn\\
&+&\left.2p_{r\alpha}\dot{\xi}^\alpha-\frac{7}{2}p_{rr}p_{vr}-2p_{rA}p_{r}^{\ A}\right],\nn\\
\eea
\bea
N^r=-1+p_{vr}+\xi^\alpha\partial_r p_{vr}+p_{r\alpha}\dot{\xi}_{\left(1\right)}^{\alpha}+\frac{1}{2}p_{r\alpha}p^{\alpha}_{\ v}-p_{rv}p_{rv},\nn\\
\eea
\bea
N^A&=&\frac{\Omega^{AB}}{\left(2M\right)^2}\left[p_{rB}+\xi_{\left(1\right)}^r\partial_r p_{rB}+\xi_{\left(1\right)}^CD_Cp_{rB}+p_{r\alpha}D_B\xi^\alpha\right.\nn\\
&+&\left(2M\right)^{-1}\xi_Bp_{rv}-p_{rA}p_{vr}+\frac{1}{2}p_{B\alpha}p^{\alpha}_{\ r}-p_{BC}p^{C}_{\ r}\nn\\
&-&2\left(2M\right)^{-1}p_{rB}\xi_{\left(1\right)}^r-\left.p_{r}^{\ C}\left[D_B\xi_{\left(1\right)C}+D_C\xi_{\left(1\right)B}\right]\right].\nn\\
\eea
In these expressions, all perturbations and their derivatives are evaluated at $x^\alpha=\left(v,2M,\theta^A\right)$. Note that $\xi^\alpha_{\left(2\right)}$ does not appear in the expressions of the basis vectors, when evaluated in terms of the intrinsic coordinates.
\section{Horizon's geometry}

In this section, we derive the form of horizon's intrinsic geometry in perturbed spacetime, characterized by the induced metric
\be
\gamma_{AB}=g_{\alpha \beta} e^{\alpha}_A e^{\beta}_B
.\ee
This metric, expressed in terms of the coordinates defined on the horizon, describes the congruence of the generators. The prescription for computing this metric is to find the corresponding first and second order Lagrangian changes to the background metric, and add it to $\gamma^{\left(0\right)}_{AB}=\left(2M\right)^2\Omega_{AB}$. The schematic form of the perturbed horizon metric up to the second order is
\bea
\phi^{*}\gamma_{AB}&=&\gamma^{\left(0\right)}_{AB}+\gamma^{\left(1\right)}_{AB}\left[\xi_{\left(1\right)}\right]+\gamma^{\left(2\right)}_{AB}\left[\xi_{\left(1\right)}\right]+\gamma^{\left(1\right)}_{AB}\left[\xi_{\left(2\right)}\right].\nn\label{gamma}\\
\eea
The first contribution, $\gamma^{\left(1\right)}_{AB}\left[\xi_{\left(1\right)}\right]$, is the linear perturbation created by the first order displacement $\xi_{\left(1\right)}$. The remaining terms belong to the second order perturbation. The second contribution, $\gamma^{\left(2\right)}_{AB}\left[\xi_{\left(1\right)}\right]$, involves bilinear perturbations involving $pp$ terms; while the last term, $\gamma^{\left(1\right)}_{AB}\left[\xi_{\left(2\right)}\right]$, is purely generated from the second order displacement $\xi_{\left(2\right)}$. Different terms in this relation are given by
\bes\bea
\gamma^{\left(1\right)}_{AB}\left[\xi_{\left(1\right)}\right]&=&p_{AB}+{\cal L}_{\xi_{\left(1\right)}}\gamma_{AB}\nn\\
&=&p_{AB}+2\left(2M\right)\xi_{\left(1\right)}^r\Omega_{AB}+\left[D_B\xi_{\left(1\right)A}+D_A\xi_{\left(1\right)B}\right],\nn\\
\eea
\begin{widetext}
\bea
\gamma^{\left(2\right)}_{AB}\left[\xi_{\left(1\right)}\right]&=&\frac{1}{2}p_{A\alpha}p^{\alpha}_{\ B}+{\cal L}_{\xi_{\left(1\right)}}p_{AB}+\frac{1}{2}{\cal L}^{\left(2\right)}_{\xi_{\left(1\right)}}\gamma_{AB}\nn\\
&=&\frac{1}{2}p_{A\alpha}p^{\alpha}_{\ B}+\xi_{\left(1\right)}^{r}D_{r}p_{AB}+\xi_{\left(1\right)}^{C}D_{C}p_{AB}+\left[p_{CB}D_A\xi_{\left(1\right)}^{C}+p_{Ar}D_B\xi_{\left(1\right)}^r-\left(2M\right)^{-1}P_{A}^{\ r}\xi_{\left(1\right)B}+A\leftrightarrow B\right]\nn\\
&+&\left(\xi_{\left(1\right)}^r\right)^{2}\Omega_{AB}+\left(2M\right)\Omega_{AB}\xi_{\left(1\right)}^{C}D_{C}\xi_{\left(1\right)}^r+\frac{1}{2}\left[\left(2M\right)^{-1}\xi_{\left(1\right)B}\left(D_A\xi_{\left(1\right)}^r\right)+\xi_{\left(1\right)}^{C}D_C\left(D_A\xi_{\left(1\right)B}\right)\right.\nn\\
&+&\left.\left(D_A\xi_{\left(1\right)}^C\right)\left[D_{C}\xi_{\left(1\right)B}+ D_{B}\xi_{\left(1\right)C}\right]+A\leftrightarrow B\right],
\eea
\bea
\gamma^{\left(1\right)}_{AB}\left[\xi_{\left(2\right)}\right]=\frac{1}{2}{\cal L}_{\xi_{\left(2\right)}}\gamma_{AB}=\frac{1}{2}\left(2M\right)\xi_{\left(2\right)}^r\Omega_{AB}+\frac{1}{2}\left(D_{B}\xi_{\left(2\right)A}+D_{A}\xi_{\left(2\right)B}\right).
\eea\end{widetext}
\ees
\subsection *{Determinant of the horizon metric and the expansion scalar}
The horizon metric of the previous section can be expressed as $\gamma_{AB}=\left(2M\right)^2\Omega_{AB}+P_{AB}$. Square root of the metric determinant up to the second order is given by \cite{Poisson+ Vlasov}
\be
\sqrt{\gamma}=\left(2M\right)^2\sin\theta\left(1+\frac{1}{2}\varepsilon+\frac{1}{8}\varepsilon^2-\frac{1}{4}\varepsilon_{A}^{\ B}\varepsilon_{B}^{\ A}\right),
\ee
where $\varepsilon_{A}^{\ B}=\frac{1}{\left(2M\right)^2}\Omega^{BC}P_{AC}$ is a second rank tensor defined on unit two sphere and $\varepsilon$ is the trace of $\varepsilon_{A}^{\ B}$. 
The expansion scalar can be computed as $\Theta=\frac{1}{2\gamma}{\dot\gamma}$. Before evaluating these terms, we introduce $\varepsilon_{A}^{\ B}$ in a schematic form
\be
\varepsilon^{A}_{\ B}=\frac{1}{\left(2M\right)^2}\Omega^{AC}\left(P_{CB}^{\left(1\right)}+P_{CB}^{\left(2\right)}\right),
\ee
where $P_{CB}^{\left(1\right)}$ and $P_{CB}^{\left(2\right)}$ are the first and second orders of perturbation that appears in  (\ref{gamma}), and $P^{\left(1\right)}$ and $P^{\left(2\right)}$ are their corresponding traces; so we have $\varepsilon=P^{\left(1\right)}+P^{\left(2\right)}$ and $\varepsilon^2=\left(P^{\left(1\right)}\right)^2+2P^{\left(2\right)}$. From these relations that the expansion scalar has the following form 
\be
\Theta=\frac{1}{2\gamma}{\dot\gamma}=\frac{1}{2}\dot{P}^{\left(1\right)}+\frac{3}{4}\dot{P}^{\left(2\right)}-\frac{1}{2}P_{AB}^{\left(1\right)}P^{AB}_{\left(1\right)}.
\ee
By evaluating these terms one will get
\begin{widetext} 
\bea
\Theta &=&\left[\frac{1}{2}\dot{p}^{A}_{\ A}+2\left(2M\right)^{-1}\dot{\xi}^{r}+D_A\dot{\xi}^{A}\right]+\frac{3}{4}\left[\dot{\xi}^r\partial_rp^{A}_{\ A}+\dot{\xi}^A D_{A}p^{B}_{\ B}+{\xi}^r\partial_r\dot{p}^{A}_{\ A}+{\xi}^A D_{A}\dot{p}^{B}_{\ B}+2\dot{p}_{r}^{\ A}D_A\xi^r\right.\nn\\
&+&2p_{r}^{\ A}D_A\dot{\xi}^r-2\left(2M\right)^{-1}\left(\dot{p}_A^{\ \ r}\xi^A+p_A^{\ \ r}\dot{\xi}^A\right)+3\left(2M\right)^{-1}\left(\dot{\xi}^C\left(D_C\xi^r\right)+{\xi}^C\left(D_C\dot{\xi}^r\right)\right)+\dot{\xi}^{C}D_C\left(D_A\xi^{A}\right)\nn\\
&+&{\xi}^{C}D_C\left(D_A\dot{\xi}^A\right)+\frac{1}{2}\dot{p}_{A\alpha}p^{A\alpha}+D_A\dot{\xi}^C\left[D_C\xi^A+D^A\xi_C\right]+\left.\left(2M\right)^{-1}\dot{\xi}^r_{\left(2\right)}+D_A\dot{\xi}^A_{\left(2\right)}\right]+\dot{p}^A_{\ B}D_A\xi^B+{p}^A_{\ B}D_A\dot{\xi}^B\nn\\
&+&\left(2M\right)^{-2}\xi^r\dot{\xi}^r-\left(2M\right)^{-1}\left(\dot{\xi}^r D_A\xi^A+\xi^rD_A\dot{\xi}^A\right)-\left(2M\right)^{-1}\left(\dot{p}^{A}_{\ A}{\xi}^r+p^{A}_{\ A}\dot{\xi}^r\right)-\frac{1}{2}\dot{p}_{AB}{p}^{AB}.
\label{Theta}\eea 
\end{widetext} 
All $\xi^\alpha$s in equation (\ref{Theta}) are $\xi^\alpha_{\left(1\right)}$, except in two terms where otherwise is specified. The next steps in this calculation are as follows: first expressing the displacement vectors in terms of the metric perturbation (with the help of equations (\ref{first order kk})-(\ref{first order ke}) and (\ref{dot xi r2})-(\ref{dot xi A2})), then substituting the perturbed metric obtained by integrating the perturbation equations in the local neighborhood of the event horizon. The first part of this procedure is lengthy but straightforward and results in a gauge invariant expression for the observable $\Theta$, valid in any tidal deformation. For the next step, however, perturbation solution in a special case must be incorporated. Exploiting the gauge invariance of this quantity, one can work in "Killing gauge" defined by $p_{\alpha\beta}t^\beta=0$, where $t^\alpha$ is the timelike killing field in this spacetime. In Eddington-Finklestein coordinates, this translates to $p_{v\alpha}=0$ and equations (\ref{first order xi}) and (\ref{second order xi}) imply that
\be
\xi_{\left(1\right)}^\alpha=\xi_{\left(2\right)}^\alpha=0\ \ \ \ {\rm at}\  r=2M.
\ee
This gives the simplified expression as
\be
\Theta=-\frac{1}{8}\dot{p}_{AB}p^{AB}.
\ee
Let specialize to the situation when the metric of the tidally deformed black hole is obtained by integrating the vacuum field equations, in the local neighborhood of the black hole. Using the decomposition of the metric perturbation in tensorial harmonics  and gauge invariant master functions $\Psi_{\rm even}$ and $\Psi_{\rm odd}$ as defined in \cite{Poisson+Martel}, the expansion scalar has the form of
\bea
\Theta&=&-\frac{1}{8\left(2M\right)^2}\Omega^{AC}\Omega^{BD}\left[\dot{\Psi}_{\rm even}\Psi_{\rm even}Y_{AB}Y_{CD}\right.\nn\\
&-&\left(\dot{\Psi}_{\rm even}\Psi_{\rm odd}+\Psi_{\rm even}\dot{\Psi}_{\rm odd}\right)Y_{AB}X_{CD}\nn\\
&+&\left.\dot{\Psi}_{\rm odd}\Psi_{\rm odd}X_{AB}X_{CD}\right].
\label{the last Theta}\eea

We consider a situation in which the tidal perturbation is switched off at times larger than $\nu=\nu_1$; so that the spacetime has the Schwarzchild metric when $\nu>\nu_1$ with the same mass parameter $M$ as the unperturbed black hole. The generally nonzero expression found for $\Theta$ implies that the horizon is not always identified by $r=2M$. By tracing null rays backward in time, we conclude that the hypersurface $r=2M$ does not extend smoothly to a null hypersurface at times $\nu<\nu_1$. In other words, the generally nonzero expression found for $\Theta$ has led us to think of a transition between a null to spacelike hypersurface. To see this, it suffices that nonstationary perturbations act \textit{and} second order perturbation theory applies. This is what we call it "nonlinear response of the black hole".

The symmetric trace-free part of the projection of $k_{\alpha;\beta}$ onto the surface spanned by $e^\alpha _A$ is defined by $\sigma_{AB}=\frac{1}{2}\left[\dot{\gamma}_{AB}-\Theta\gamma_{AB}\right]$. In "Killing gauge", it is given by
\be
\sigma_{AB}=\frac{1}{2}\left[\dot{p}_{AB}+\dot{p}_{AC}p^{C}_{\ B}+\frac{1}{8}\left(2M\right)^2\Omega_{AB}\dot{p}_{CD}p^{CD}\right].
\ee

A quasi-local analysis of the perturbed black hole suggests the characterization of a black hole as a spacetime trapped region.
The infinitesimal variations of the area of the emitted light front from the closed spatial surface spanned by basis vectors $e^{\alpha}_{A}$ along the outer directions is given by $\Theta^{(k)}:=\Theta$, while that of the ingoing expansions is characterized by $\Theta^{(N)}$, defined by $e_{\alpha}^{\ A} e^{\beta}_A N^{\alpha}_{\ ;\beta}$. With a lengthy but straightforward calculation one can show that the associated expression in "Killing gauge" is given by
\be
\Theta^{(N)}=-\frac{1}{2}\partial _r{p}^{A}_{\ A}+\frac{1}{2}p_{r}^{A}D_{A}P^{B}_{\ B}+\frac{1}{8}p^{AB}\partial _r{p}_{AB}
.\ee
Contrary to $\Theta^{(k)}$, the scalar $\Theta^{(N)}$ is gauge dependent.
The sign of $\Theta^{(k)}$ and $\Theta^{(N)}$ determine the characterization of closed spacelike 2-surfaces, located inside the black hole, in spacetime evolution. Those on which $\Theta^{(k)}\Theta^{(N)}>0$, are called trapped surfaces and the notion of marginally outer trapped surface is given by $\Theta^{(k)}=0$ and $\Theta^{(N)}<0$. These intrinsically quasi-local surfaces form spacelike worldtubes of the trapping horizon, with no reference to asymptotic quantities. The trapping horizon is a future outer type, if $\Theta^{(N)}<0$ and a displacement along $N^{\alpha}$ (ingoing direction) takes us to the trapping region, i.e. ${\cal L}_N \Theta^{(k)}<0$; so that the trapping horizon should be outer. 

Eq.(\ref{the last Theta}) shows that the cross sections of the perturbed surface, generated by the vector field $\xi$, are not marginally outer trapped surfaces. Surprisingly, this non-conformity does \textit{not} show up at first order. On the other hand, since $\sigma_{AB}\neq 0$, such surfaces can not evolve into a nonexpanding horizon \cite{Dreyer}. 

In studying a perturbed black hole, the seminal notion of a future outer trapping horizon, ${\cal FOTH}$, plays a crucial role as a potential black hole boundary. The qualitative and quantitative aspects of a dynamical trapping horizon are generically given by a PDE system in an Initial Value Problem approach \cite{Booth+Jaramillo}. The evolution for a given marginally outer trapping surface into a dynamical ${\cal FOTH}$ containing this surface is possible provided that ${\cal L}_N\Theta^{(k)}<0$. 
Eventually, the evolution may result in the formation of a new horizon around the old one (or what we call the black hole response), in either smoothly or in a discontinuous jumping manner, where it ceases to emit gravitational waves. 
\section{Conclusion}
In this paper, the second order tidal perturbation of a nonrotating black hole is studied and a second order invariant observable in this theory is presented. This is a scalar quantity, denoted by $\Theta$, describing the growth in area of any cross section $v$=constant of outgoing light front on the trapping horizon.
 The main result of the paper is given in equation (\ref{the last Theta}). This scalar starts at order $1/\left({{\cal T}{\cal R}^4}\right)$, where the scales $\cal R$ and $\cal T$ specify the radius of local curvature and timescale of the changes in the tidal environment, respectively. This result shows that the horizon boundary in a tidally deformed black hole may be a spacelike surface and the surface $r=2M$ does not extend smoothly to the times before settle down. This behaviour does not show up at linear order. If the tidal processes are slow, in the sense that their characteristic timescales are slow compared to the black hole mass, the expansion scalar vanishes. This is an exact result up to first nonlinear order and can be compared to that given in reference \cite{Poisson+ Vlasov}. This surprising result implies that in studying these systems, $\Theta$ is a 3rd order invariant scalar. Note that this does {\it not} generally implies that $\Theta$ is a third order gauge invariant observable. Under restrictive assumptions, however, like static or slowly varying tides, this is a correct statement. This study can be considered as a starting point in better understanding of the geometry of the trapped region in ringdown phase of a binary evolution. Although the deformations may well not related to the presence of a tidal disturber and could therefore be considered in vacuum spacetimes.
\section *{Acknowledgments} 
The author would like to thank the University of Tehran for supporting this project under a research grant 
provided by the university research council.


\begin{thebibliography}{100}

\bibitem{Vega+Poisson}
I. Vega, E. Poisson and R. Massey, Class. Quantum Grav. {\bf 28}, 175006, (2011). 

\bibitem{Flanagan+Hinderer}
M. Shibata and K. Taniguchi, Phys. Rev. D{\bf 77}, 084015,(2008); E. E. Flanagan and T. Hinderer, Phys. Rev. D {\bf 77}, 021502, (2008); T.  Hinderer, B. D. Lackey, R. N. Lang and J. S. Read, Phys. Rev.D {\bf 81}, 123016, (2010).    
 
\bibitem{Damour and Nagar}
T. Damour and A. Nagar, Phys. Rev. D{\bf 80}, 024019, (2008); T. Binnington and E. Poisson, Phys. Rev. D{\bf 80}, 084018, (2009).


\bibitem{Stewart and Walker}
Sachs, H. {\it In Relativity, groups and topology} (eds B. IDeWitt and C. De Witt). New York: Gordon and Breach, I964; J. M. Stewart and M. Walker, Proc. R. Soc. London A. {\bf 341}, 49 (1974).

\bibitem{Bruni and Sonego}
M. Bruni and S. Sonego, Class. Quant. Grav. {\bf 16}, L29, (1999). 

\bibitem{Schutz and Sorkin}
B. F. Schutz and R. Sorkin, Ann. Phys. {\bf 107}, 1,(1977); J. L. Friedman, Commun. math. Phys. {\bf 62}, 247 (1978).

\bibitem{Friedman and Schutz}
J. L. Friedman and B. F. Schutz, Astrophysics J. {\bf 221}, 937 (1978).

\bibitem{Poisson}
E. Poisson, Phys. Rev. D {\bf 70}, 084044  (2004).
 
\bibitem{Bruni et al}
 M. Bruni, S. Matarrese, S. Mollerach and S Sonego, Class. Quant. Grav. {\bf 14}, 2585 (1997).  

\bibitem{Poisson+ Vlasov}
E. Poisson and I. Vlasov, Phys. Rev. D, {\bf 81}, 024029, (2010).

\bibitem{Poisson+Martel}
K. Martel and E. Poisson , Phys. Rev. D, {\bf 71}, 104003, (2005).

\bibitem{Dreyer}
O. Dreyer, B. Krishnan, D. Shoemaker and E. Schnetter, Phys. Rev. D, {\bf 67}, 024018 (2003).

\bibitem{Booth+Jaramillo}
A. Ashtekar and B. Krishnan, Living Rev. Relativity, 7, (2004), 10. [Online Article]: cited [28 January 2008],
http://www.livingreviews.org/lrr-2004-10; I. Booth,  Can. J. Phys, {\bf 83}, 1073,(2005); J. L. Jaramillo,  Notes prepared for the course at the 2011 Shanghai Asia-Pacific School and Workshop on Gravitation (Shanghai Normal University, February 10-14, 2011). 

 
\end{thebibliography}
\end{document}